\documentclass{jps-cp}
\usepackage{txfonts} 
\usepackage{bm}
\newcommand{\Slash}[1]{{\ooalign{\hfil/\hfil\crcr$#1$}}}

\title{Operator Relations for Gravitational Form Factors}

\author{Kazuhiro \textsc{Tanaka}}

\inst{Department of Physics, Juntendo University, Inzai,
  Chiba 270-1695, Japan and\\
J-PARC Branch, KEK Theory Center, Institute of
  Particle and Nuclear Studies, KEK, 203-1, Shirakata, Tokai, Ibaraki,
  319-1106, Japan
}

\email{kztanaka@juntendo.ac.jp}

\abst{The form factors for the hadron matrix element of the QCD energy-momentum tensor not only describe the coupling of the hadron with a graviton as the ``gravitational form factors'', but also serve as unique quantities for describing the shape inside the hadron reflecting dynamics of quarks and gluons, such as the internal shear forces acting on the quarks/gluons and their pressure distributions. We consider  
the gravitational form factors for a hadron, in particular, for a (pseudo)scalar hadron and for the nucleon. We derive and clarify the relations satisfied by the gravitational form factors as direct consequences of the symmetries and the equations of motion in QCD, and connections to the generalized parton distributions. Our results reveal that the gravitational form factors are related to the higher-twist quark-gluon correlation effects inside the hadrons and also to QCD trace anomaly.}

\kword{gravitational form factors, energy-momentum tensor, QCD, trace anomaly}

\begin{document}
\vspace*{-1cm}
\begin{flushright}
J-PARC TH-0164
\end{flushright}

\maketitle

\section{Introduction}
The (Belinfante-improved) energy-momentum tensor in QCD is a symmetric second rank tensor expressed by the sum 
of the gauge-invariant quark part and gluon part as~\cite{Polyakov:2018zvc,Tanaka:2018wea}
\begin{equation}
T^{\mu\nu} = \frac{1}{2}\bar{\psi} \gamma^{(\mu} i\overleftrightarrow{D}^{\nu )} \psi+ \left(F_a^{\mu\rho}{{F_a}_\rho}^\nu+\frac{g^{\mu \nu}}{4}F_a^{\lambda\rho}{F_a}_{\lambda\rho}\right)\equiv   T^{\mu\nu}_q+ T^{\mu\nu}_g \ ,
\label{tqg}
\end{equation}
where $D^\mu=\partial^\mu-ig A^\mu$, $R^{(\mu}S^{\nu)}\equiv \left(R^\mu S^\nu+R^\nu S^\mu\right)/2$, and 
we have neglected the ghost and gauge fixing terms, as well as the terms that vanish by the use of the QCD equations of motion,  as they do not affect our final results. 
The matrix element of  each term in (\ref{tqg})
sandwiched between the nucleon states $|p\rangle$ and $|p'\rangle$, with $|p\rangle$ being associated 
with momentum $p$, mass $m_N$ and the spinor $u(p)$, as
\begin{equation}
 \langle p'|T_{q,g}^{\mu\nu}|p\rangle = \bar{u}(p')\Bigl[A_{q,g} (t)\gamma^{(\mu}\bar P^{\nu)}
 +B_{q,g}(t)\frac{\bar P^{(\mu}i\sigma^{\nu)\alpha}\Delta_\alpha}{2m_N}
 + D_{q,g}(t)\frac{\Delta^\mu\Delta^\nu -g^{\mu\nu}t}{4m_N} + \bar{C}_{q,g}(t)m_Ng^{\mu\nu}\Bigr] u(p)
\label{para}
\end{equation}
is described by the gravitational form factors,
 $A_{q,g} (t), B_{q,g}(t), D_{q,g}(t)$, and $\bar{C}_{q,g}(t)$, 
where
 $\Delta=p'-p$, $\bar{P}=\left(p+p'\right)/2$, 
$t=\Delta^2$, and $\bar{P}^2= m_N^2 -t/4$.
Those gravitational form factors of hadrons have received considerable attention recently~\cite{Burkert:2018bqq,Polyakov:2018zvc}.
Although it is impractical to detect 
them as
the coupling of hadrons with a graviton,
it is now realistic to determine the nucleon's gravitational form factors 
through behaviors 
of the generalized parton distributions (GPDs),
obtained by experiments like deeply virtual Compton scattering,
deeply virtual meson production,
etc~\cite{Burkert:2018bqq};
this relies on relations,
\begin{equation}
\int_{-1}^{1} dxx H^{q} (x, \eta,t) =A_q (t)+  \eta^2D_q(t)\ ,\;\;\;\;\;\;
\int_{-1}^{1} dxxE^{q} (x, \eta,t) =B_q(t)- \eta^2D_q(t)\ ,
\label{abd}
\end{equation}
for the moments of the GPDs; here, the relevant GPDs are defined as usual
(see, e.g., \cite{Belitsky:2005qn,Sawada:2016mao}):
\begin{equation}
\int \frac{dz^ - }{2\pi } e^{ix\bar P^+z^-}\langle p'| \bar \psi( - \frac{z}{2}) \gamma ^ + \psi(\frac{z}{2}) |p\rangle\Big |_{z^+ = \bm{z}_\perp =0}  = \frac{1}{\bar P^ + }\bar u(p')\left[ H^q(x,\eta ,t) \gamma ^ +  
+ E^q(x,\eta ,t) \frac{i\sigma ^{ + \alpha }\Delta_\alpha }{2m_N}  \right]u(p)\ ,
\end{equation}
with the scaling variable $x$ and the skewness parameter $\eta=-\Delta^+/(2 \bar{P}^+)$, 
where $a^\pm=(a^0 \pm a^3)/\sqrt{2}$ denote the
plus/minus light-cone components of a four-vector $a^\mu$, such that $x$
and $2\eta$ are the light-cone momentum fractions of the average
momentum and momentum transfer for the relevant quarks, respectively,
to the average momentum $\bar P$ of the parent nucleon.
It has been demonstrated~\cite{Burkert:2018bqq} that 
combining (\ref{abd}) with the behaviors 
of the GPDs obtained by the above-mentioned experiments
allows us to determine, in particular, 
$D_{q,g}(t)$ of (\ref{para}),
which are related to 
the D term, $D\equiv D_q(0)+D_g(0)$, and 
are
considered as the last unknown fundamental hadron characteristic
determining the spatial deformations as well as defining
the mechanical properties of hadrons~\cite{Polyakov:2018zvc}.
Thus, the investigation of the gravitational form factors is a hot topic,
and an urgent task from the theory side is to clarify a maximal set of
(exact and approximate) relations satisfied by the gravitational form factors in (\ref{para}).

\section{Constraints from symmetries and equations of motion in QCD}

Based on parity (P) invariance combined with time-reversal (T) invariance, we can show 
$\langle p'|T_{q,g}^{\mu \nu}|p\rangle$
$=\langle p'|
T_{q,g}^{\mu \nu}|p\rangle^\ast
=\langle p|
T_{q,g}^{\mu \nu}|p'\rangle$;
therefore, 
the gravitational form factors are real quantities, and
(\ref{para}) is the most general form satisfying the symmetry constraints.
We also note that the divergenceless property of (\ref{tqg}), $\partial _\mu T^{\mu\nu}(x)=0$, implies
$\bar{C}_q(t)+\bar{C}_g(t)=0$.
Furthermore, in the forward limit, $\Delta \to 0$, $t\to 0$,
we have the sum rules,  
$A_q(0)+A_g(0)=1$,
due to the fact that
$\langle p|T^{\mu\nu}|p\rangle =2p^\mu p^\nu$ holds for (\ref{para}),
representing the total energy-momentum,
and $\left(A_q(0)+B_q(0) +A_g(0) +B_g(0)\right)/2=1/2$, due to the fact 
that $\langle p|J^i|p\rangle/\langle p|p\rangle =\left(1/2\right)S^i$ holds with $S^i$ the nucleon's spin vector, $J^i = \epsilon^{ijk}\int
dx^-  d^2 x_\perp M^{ + jk}/2$, and ${M^{\mu \rho \sigma }} = {x^\rho }{T^{\mu \sigma }} - {x^\sigma }{T^{\mu \rho }}$,
representing the total angular momentum.

Exact manipulations for the divergence of  (\ref{tqg}), keeping all the contributions of quark masses, e.g., 
\begin{equation}
-i\partial_\nu \left\{-\bar{\psi} i\overleftarrow{D}^\nu\gamma^{\mu}\psi\right\}
=-\frac{1}{2}\bar{\psi}  g F_{\nu \alpha}\sigma^{\nu \alpha}\gamma^{\mu}\psi+ m^2\bar{\psi}  \gamma^{\mu}\psi
+ \bar{\psi} i\overleftarrow{D}^\nu i\overrightarrow{D}_\nu \gamma^{\mu}\psi+{\rm EOM}
\ ,
\end{equation}
where $\overleftarrow{D}^\mu=\overleftarrow{\partial}^\mu+igA^\mu$
and the ``EOM'' denotes the operators that vanish by the use of the equations of motion,
eventually give the significantly compact operator identities~\cite{Tanaka:2018wea},
\begin{equation}
\partial _\nu T_q^{\mu \nu } =  - \bar \psi g{F^{\mu \nu }}{\gamma _\nu }\psi\ ,\;\;\;\;\;\;\;\;\;\;\;\;\;\;\;\;
\partial _\nu T_g^{\mu \nu } =  - F_a^{\mu \nu }D_{ab}^\rho F_{\rho \nu }^b\ ,
\label{eoi}
\end{equation}
up to the terms that vanish by 
the QCD equations of motion, 
$\left(i\Slash{D}-m\right)\psi=0$, and $[D_\mu ,  F^{\mu \nu}]=-t^a g\bar{\psi}t^a \gamma^\nu \psi$.
Taking the matrix element of (\ref{eoi}), we find
the exact relations:
\begin{equation}
 \Delta^\mu\bar{u}(p')u(p) m_N\bar{C}_q(t)=\langle p'|\bar{\psi}ig F^{\mu \nu}\gamma_\nu \psi|p \rangle\ ,
\;\;\;\;\;\;\;\;
\Delta^\mu \bar{u}(p')u(p) m_N\bar{C}_g(t) 
=\langle p'|F_a^{\mu \nu} iD^\rho_{ab}  F^b_{\rho \nu}|p\rangle\ ,
\label{barc}
\end{equation}
which show that $\bar{C}_{q}(t)=-\bar{C}_{g}(t)$ is related to the quark-gluon interactions of twist four and higher.

\section{Unravelling in the light-cone gauge fixing}

Exact gauge-invariant manipulations discussed above, utilizing symmetries and the equations of motion, 
allow us to obtain several constraints on the gravitational form factors in (\ref{para}) as well as
the explicit operator content of the form factor $\bar{C}_{q,g}(t)$
as (\ref{barc}). However, this approach is not helpful for revealing direct information on 
the form factor corresponding to the D term.
To try to assess the operator content of $D_q(t)$ particular to the off-forward matrix element,
we employ gauge fixing to allow us to treat each term of the covariant derivative in (\ref{tqg}) separately and identify the physical degrees 
of freedom~\cite{Tanaka:2018wea}.
We take the light-cone gauge $n_\mu A^\mu=0$ with a lightlike vector $n^\mu=g^\mu_-n^-$, 
anticipating manipulations linked with the partonic interpretations 
appropriate in the infinite momentum frame;
then, 
the gluon field in
the matrix element of the quark part of the energy-momentum tensor of (\ref{tqg}), 
\begin{eqnarray}
\langle p'|T_q^{\mu \nu}|p\rangle
=
\frac{1}{4}\langle p'|\bar{\psi}\left( -i\overleftarrow{\partial}^\mu + i\overrightarrow{\partial}^\mu+2gA^\mu\right)\gamma^{\nu} \psi
|p\rangle + \left( \mu \leftrightarrow  \nu \right)\ ,
\label{hw0}
\end{eqnarray}
can be expressed by the field strength tensor as
$A^\mu(\lambda n) =\frac{1}{2}\int d\lambda'\ {\rm sgn} (\lambda'-\lambda) F^{\mu \alpha}(\lambda' n)n_\alpha$,
with ${\rm sgn}(\lambda)=\theta(\lambda)-\theta(-\lambda)$,
assuming the antiperiodic boundary condition for $A^\mu(y)$ at $|y^-|\to \infty$,
while the derivative terms are handled, using the Heisenberg equations for the quark and antiquark field operators, as
\begin{equation}
\langle p'|\bar{\psi}\left( -i\overleftarrow{\partial}^\mu + i\overrightarrow{\partial}^\mu\right)\gamma^{\nu} \psi
|p\rangle
=\langle p'|\left(   \bar{\psi}\gamma^{\nu} \left[ \psi, \hat{\cal P}^\mu\right] -\left[ \bar{\psi}, \hat{\cal P}^\mu\right]\gamma^{\nu}\psi \right)
|p\rangle
=2\bar{P}^\mu\langle p'|    \bar{\psi}\gamma^{\nu}  \psi|p\rangle
-2\langle p'|\bar{\psi} \hat{\cal P}^\mu\gamma^{\nu}\psi|p\rangle\ ,
\label{remove}
\end{equation}
with the 4-momentum operator, $\hat{\cal P}^\mu = \int dx^- d^2x_\perp\,  T^{\mu +}(x)$, in the light-cone quantization of QCD.
The second term is evaluated by inserting a 
complete set of the light-cone Fock states, $\sum_r|p_r\rangle \langle p_r | =1$, as
\begin{equation}
\langle p'|\bar{\psi} \hat{\cal P}^\mu\gamma^{\nu}\psi|p\rangle
=\sum_r p_r^\mu\langle p'|\bar{\psi} |p_r\rangle \langle p_r | \gamma^{\nu}\psi|p\rangle
= \langle p_r^\mu \rangle \langle p'|    \bar{\psi}\gamma^{\nu}  \psi|p\rangle
=\left(p^\mu-\langle k_q^\mu\rangle\right)\langle p'|    \bar{\psi}\gamma^{\nu} \psi|p\rangle
\ ,
\label{averagev00}
\end{equation}
where $\langle p_r\rangle$ denotes the value of $p_r$ averaged over intermediate states,
and $p_r$ equals $p-k_q$
where $k_q$ denotes the 4-momentum of the
quark removed from the initial state $|p\rangle$ by the action of the field operator $\psi$.
Noting that, $k_q^\mu  =x \bar{P}^\mu-\Delta^\mu/2$, in the parton language relevant for the GPD formulation,
which is considered to be appropriate 
for $k_q^\mu$ with $\mu=+$ and $\perp$ when taking (the light-cone quantization in) the light-cone gauge, and assuming that this identification is accurate in the averaging 
for $\langle k_q^\mu \rangle$
in (\ref{averagev00}), 
we obtain 
$p^\mu-\langle  k_q^\mu\rangle\simeq p^\mu- \left\langle x \bar{P}^\mu-\Delta^\mu/2\right\rangle
= \bar{P}^\mu-\langle x\rangle \bar{P}^\mu$,
using $p=\bar{P}-\Delta/2$.
Combining this with the above formula (\ref{averagev00})
for $\mu=+, \perp$, and 
substituting the result into (\ref{remove}), (\ref{hw0}), we obtain~\cite{Tanaka:2018wea}
\begin{equation}
\langle p'|T_q^{\mu \nu}|p\rangle\simeq
\bar u(p')\left[ \langle x \rangle F_1^q(t) \bar{P}^{(\mu}  \gamma^{\nu )} +\langle x \rangle  F_2^q(t)
  \frac{\bar{P}^{(\mu} i\sigma^{\nu ) \alpha}\Delta_\alpha}{2m_N}\right]u(p)
+ \frac{1}{2}\langle p'|\left( \bar{\psi}gA^\mu\gamma^{\nu}\psi+ \bar{\psi}gA^\nu\gamma^{\mu}\psi\right) 
|p\rangle\ ,
\label{hw}
\end{equation}
where $F_{1,2}^q (t)$ are the usual Dirac and Pauli form factors for the nucleon.
Comparing this with (\ref{para}),
\begin{eqnarray}
&& A_q(t) +\eta^2D_{q} (t)
\simeq
 \langle x \rangle F_1^q(t)\ , \;\;\;\;\;\;\;\;\;\;\;\;
B_q (t)-\eta^2D_{q} (t)\simeq
 \langle x \rangle F_2^q(t)\ ,
\label{paranovelr}\\
&&-\frac{\eta\Delta_\perp^\mu\bar{u}(p')u(p)}{m_N} D_{q}(t)
\simeq
\int_{-\infty}^\infty d\lambda\ \frac{{\rm sgn} (\lambda)}{2} 
n_\alpha\langle p'| gF_a^{\mu \alpha}(\lambda n)
\bar{\psi}(0)t^a\Slash{n} \psi(0)
|p\rangle
\label{last}
\end{eqnarray}
are obtained;
if one makes the replacement, $x \to \langle x \rangle$, under the integration in the LHS in (\ref{abd}),
we obtain the result formally similar to (\ref{paranovelr}).
On the other hand, a similar logic is not applicable to (\ref{last});
this fact suggests the nontrivial nature of  (\ref{last}), which
implies that $D_q(t)$ corresponds to an integral of
the twist-three quark-gluon correlation, in particular, 
the correlation between the quark color-current density and the gluon field strength.
Remarkably, the results (\ref{paranovelr}), (\ref{last}) indicate that the large $t$ behavior of 
$A_q(t)$ should be similar as the quark counting~\cite{Belitsky:2005qn} for the vector form factor $F_1^q(t)$,
while $D_q(t)$ should receive an additional $1/t$ suppression due to the quark counting for the three-body Fock state: because $F_1^q(t\to \infty) \sim 1/t^2$, we expect the large $t$ behavior to be
$A_q(t) \sim 1/t^2$,  and $D_q(t) \sim 1/t^3$.

\section{Gravitational form factors for the pion}

Some of our results discussed above are immediately extended to the corresponding relations for the  
gravitational form factors for a spin-0 hadron such as a pion~\cite{Tanaka:2018wea}. Denoting a spin-0 hadron state with 4-momentum $p$ and mass $m_h$
as $|h(p)\rangle$, the corresponding gravitational form factors are defined as
\begin{equation}
\langle h(p')|T_{q}^{\mu \nu}|h(p)\rangle
=
\frac{1}{2}{\Theta}_{1q}(t)\left(tg^{\mu\nu}- \Delta^\mu \Delta^\nu\right)+ \frac{1}{2}{\Theta}_{2q}(t) \bar{P}^\mu\bar{P}^\nu+ \Lambda^2\bar{C}^h_q(t)g^{\mu \nu}\ ,
\label{tmunumatred}
\end{equation}
where $\Lambda$ denotes a nonperturbative mass scale in QCD,
and the matrix element of the gluon part of (\ref{tqg}) is given by the similar parameterization with $q\to g$. 
The exact relations (\ref{barc}) with the substitutions, $\bar{u}(p')u(p) m_N\bar{C}_q(t)\to \Lambda^2\bar{C}^h_q(t)$
and $|p\rangle\to|h(p)\rangle$, are derived~\cite{Tanaka:2018wea}, and, similarly to (\ref{paranovelr}), (\ref{last}),
we find
\begin{equation}
\frac{{\Theta}_{2q}(t)}{2} -2\eta^2{\Theta}_{1q}(t)
\simeq 2\langle x \rangle{\cal F}_v(t)\ ,\;
2 \eta\Delta_\perp^\mu {\Theta}_{1q}(t) \simeq
\int_{-\infty}^\infty d\lambda \frac{{\rm sgn} (\lambda)}{2} 
n_\alpha\langle h( p')| gF_a^{\mu \alpha}(\lambda n)
\bar{\psi}(0)t^a\Slash{n} \psi(0)
|h(p) \rangle
\ ,
\label{lctheta1}
\end{equation}
where ${\cal F}_v(t)$ denotes the quark contribution of the vector form factor, 
$\langle h(p')|\bar{\psi}  \gamma_\mu \psi|h(p)\rangle=2 {\cal F}_v(t)\bar{P}_\mu$.
Thus, combined with ${\cal F}_v(t\to \infty) \sim 1/t$ from the dimensional counting rule,
we expect the large $t$ behavior as
${\Theta}_{2q}(t) \sim 1/t$, ${\Theta}_{1q}(t) \sim 1/t^2$~\cite{Tanaka:2018wea}.
Those results should be relevant for extracting
the behaviors of the gravitational form factors of the pion using, e.g., the Belle data on $\gamma^* \gamma  \rightarrow \pi^0 \pi^0$~\cite{Kumano:2017lhr}.

\section{Constraints from the QCD trace anomaly}

The gravitational form factors $\bar{C}_{q,g}$ of (\ref{para}) are also interesting because they are relevant to
the force distribution inside the nucleon~\cite{Polyakov:2018zvc,Polyakov:2018exb,Hatta:2018ina,Lorce:2018egm}
and the nucleon's transverse spin sum rule~\cite{Hatta:2012jm}.
In addition to the relations (\ref{barc}) useful for $\Delta\neq 0$, constraints on 
$\bar{C}_{q,g}(t=0)$ 
can be obtained by the new relations~\cite{Hatta:2018sqd} representing the decomposition of the QCD trace anomaly
into the quark and gluon contributions.

The trace of the forward limit, $\Delta \to 0$, of (\ref{para}) reads,
\begin{equation}
{g_{\mu \nu }}\langle p|T_{q,g}^{\mu \nu }|p\rangle  = 2{m_N^2}\left( {{A_{q,g}}(0) + 4{{\bar C}_{q,g}}(0)} \right)\ .
\label{trace}
\end{equation}
Here, in the LHS, ${g_{\mu \nu }}T_q^{\mu \nu } = m\bar \psi\psi$ and ${g_{\mu \nu }}T_g^{\mu \nu } = 0$, up to the EOM,
are suggested by 
(\ref{tqg}), but those do not hold; in general, the trace of a symmetric second rank tensor 
receives the anomaly contributions of ${\cal O}(\alpha_s)$ and higher, such that the trace operation and the renormalization do not commute.
We find
\begin{equation}
g_{\mu \nu }T_q^{\mu \nu }  =  m\bar \psi\psi + \frac{{{\alpha _s}}}{{4\pi }}\left( {\frac{{{n_f}}}{3}{F^2} + \frac{{4{C_F}}}{3}m\bar \psi\psi} 
\right)\ , \;\;\;\;\;\;\;\;
g_{\mu \nu }T_g^{\mu \nu }  = \frac{{{\alpha _s}}}{{4\pi }}\left( { - \frac{{11{C_A}}}{6}{F^2} + \frac{{14{C_F}}}{3}m\bar \psi\psi} \right)\ ,
\label{ano}
\end{equation}
at one loop in the (modified) minimal subtraction scheme in the dimensional regularization, 
for $n_f$ flavor and $N_c$ color with $C_F=(N_c^2-1)/(2N_c)$ and $C_A=N_c$~\cite{Hatta:2018sqd}.
These results combined with (\ref{trace}) can be used to constrain  $\bar{C}^R_{q,g}\equiv \bar{C}_{q,g}(t=0, \mu)$
as a function of the renormalization scale $\mu$; e.g., for $\bar{C}^R_q$ with a certain starting scale $\mu_0$,
we obtain ($A_q^{R0}\equiv A_q\left(t=0, \mu=\mu _0\right)$, $\beta_0\equiv(11C_A-2n_f)/3$)
\begin{equation}
\bar{C}^R_q=
- \frac{1}{4} \left( \frac{n_f}{4C_F+n_f} + \frac{2n_f}{3\beta_0}\right) + \left(\frac{2 n_f}{3 \beta _0}+1\right)\frac{\left\langle p \right| m \bar{\psi }\psi  \left| p \right\rangle}{8m_N^2}
   -\frac{C_F A_q^{R0}+\frac{n_f}{4} \left(A_q^{R0}-1\right)}{4 C_F+n_f}\left(\frac{\alpha _s\left(\mu
   \right)}{\alpha _s(\mu_0 )}\right)^{\frac{ 8 C_F+2n_f}{3 \beta_0}}\ ,
   \label{asy}
\end{equation}
where the first term predicts the asymptotic value in the chiral limit, $\bar{C}^R_q\to -0.146$ ($n_f=3$) and $\bar{C}_q^R\to -0.103$ ($n_f=2$),
as $\mu\to \infty$~\cite{Hatta:2018sqd}. The results in (\ref{ano}), (\ref{asy}) are extended to the two-loop order~\cite{Hatta:2018sqd}, and also to the three-loop order~\cite{Tanaka:2018nae}, such that the total sum of the 
corresponding formulas, $g_{\mu \nu }T_q^{\mu \nu }+ g_{\mu \nu }T_g^{\mu \nu }$, reproduces the well-known QCD trace anomaly,
$T^\mu_\mu= (\beta/2g)F^{\mu\nu}F_{\mu\nu}+ m (1+\gamma_m)\bar{\psi}\psi$.

Finally, combining 
the QCD trace anomaly with the relation, $\langle p|T^\mu_\mu|p\rangle =2m_N^2$, it is frequently argued that
the entire mass $m_N$ comes from gluons in the chiral limit;
but, (\ref{ano}) 
shows~\cite{Tanaka:2018nae} that 
the quark (gluon) loop effects make the nucleon mass 
light (heavy), with $\langle p|\left. F^2\right|_{\mu=1~\rm{GeV}}|p\rangle\simeq-8.61m_N^2$.

\smallskip
\noindent
{\bf Acknowledgments}

I thank S.~Kumano, Q.~T.~Song, Y.~Hatta and A.~Rajan for discussions and collaborations.   


\end{document}